\documentclass[10pt,twocolumn,letterpaper]{article}

\usepackage{iccv}
\usepackage{times}
\usepackage{epsfig}
\usepackage{graphicx}
\usepackage{amsmath}
\usepackage{amssymb}
\usepackage{multirow}

\usepackage[pagebackref=true,breaklinks=true,letterpaper=true,colorlinks,bookmarks=false]{hyperref}

\iccvfinalcopy % *** Uncomment this line for the final submission

\def\iccvPaperID{70} % *** Enter the ICCV Paper ID here

\ificcvfinal\fi

\begin{document}

%%%%%%%%% TITLE

\title{Dual Reconstruction with Densely Connected Residual Network \\for Single Image Super-Resolution}

\author{Chih-Chung Hsu\\
	Department of Management Information Systems \\National Pingtung University of Science and Technology\\
	No. 1, Shuefu Road, Neipu, Pingtung 91201, Taiwan\\
	{\tt\small cchsu@mail.npust.edu.tw}
	\and
	Chia-Hsiang Lin\\
	Department of Electrical Engineering, \& Institute of Computer and Communication Engineering\\ National Cheng Kung University\\
	No.1, University Road, Tainan City 701, Taiwan\\
	{\tt\small chiahsiang.steven.lin@gmail.com}
}

\maketitle
\ificcvfinal\thispagestyle{empty}\fi

%%%%%%%%% ABSTRACT
\begin{abstract}
   Deep learning-based single image super-resolution enables very fast and high-visual-quality reconstruction. 
   Recently, an enhanced super-resolution based on generative adversarial network (ESRGAN) has achieved excellent performance in terms of both qualitative and quantitative quality of the reconstructed high-resolution image. 
   In this paper, we propose to add one more shortcut between two dense-blocks, as well as add shortcut between two convolution layers inside a dense-block.
   With this simple strategy of adding more shortcuts in the proposed network, it enables a faster learning process as the gradient information can be back-propagated more easily. 
   Based on the improved ESRGAN, the dual reconstruction is proposed to learn different aspects of the super-resolved image for judiciously enhancing the quality of the reconstructed image. In practice, the super-resolution model is pre-trained solely based on pixel distance, followed by fine-tuning the parameters in the model based on adversarial loss and perceptual loss. Finally, we fuse two different models by weighted-summing their parameters to obtain the final super-resolution model. Experimental results demonstrated that the proposed method achieves excellent performance in the real-world image super-resolution challenge. We have also verified that the proposed dual reconstruction does further improve the quality of the reconstructed image in terms of both PSNR and SSIM.
\end{abstract}
%%%%%%%%% BODY TEXT
\section{Introduction}
%% Some preview...
%%==================Figure====================
\begin{figure}
	\centering
	\includegraphics[width=0.5\textwidth]{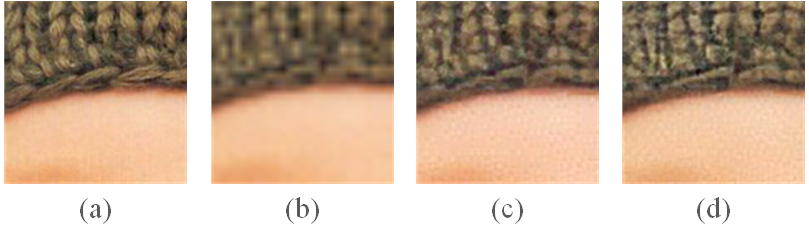}
	\caption{\small Visual quality comparison between the (a) ground truth and the super-resolved results by using (b) Bicubic interpolation, (c) ESRGAN \cite{esrgan}, and (d) proposed method. }
	\label{fig:header}
\end{figure}
%%==================Figure====================
Due to limitations in hardware design, or long distance between the object and the sensors (e.g., in satellite images), spatial resolution of the acquired images is quite often unsatisfactory.
Without relying on hardware improvement, we aim at improving the image spatial resolution computationally.
One category of image super-resolution (SR) techniques is referred to as panchromatic sharpening \cite{lee2009fast,lin2019simiregPS}, for which the camera acquires a single-band image (hence allowing very high-spatial resolution acquisition) of the same spatial region while acquiring the target imagery.
Then, the acquired single-band image, known as panchromatic image, provides essential spatial details (or high-frequency components) that are injected/fused into the target imagery, thereby computationally super-resolving the target imagery.

This kind of image fusion approach is of paramount importance in nowadays satellite remote sensing technology, and also appears in other domain applications, such as the hyperspectral super-resolution technique \cite{lin2017COCNMF} which fuses a high-resolution multispectral image with a low-resolution hyperspectral image in order to obtain a high spatial/spectral-resolution imagery---critical in military surveillance.
However, these approaches all rely on high-resolution counterpart imageries that are, however, not always available, motivating us to pursue single-image super-resolution (SISR) for facilitating the ensuing computer vision tasks \cite{huang2015single}.

Besides the well-known bicubic interpolation, several conventional SISR methods have been studies, including Lanczos resampling \cite{duchon1979lanczos} and optimization based algorithms, where the latter usually considers the SISR problem as an imaging inverse problem \cite{bertero1998introduction}.
SISR has also been widely studied in statistical machine learning area, which requires judiciously transforming suitable prior knowledge into computationally friendly regularization schemes \cite{kim2010single}, such as self-similarity \cite{Sentinel2SR} and total-variation \cite{ng2007total}.

Recently, deep learning has revealed its potential in achieving high performance SISR task. In the convolutional neural network (CNN) framework, an end-to-end sparse coding CNN (SRCNN), not requiring any engineered features, has also been proposed in \cite{dong2015image}.
Though SRCNN is a pioneering work introducing the deep learning in solving SR, it converges slowly and relies on the context of small regions. Furthermore, a very deep neural network SR (VDSR) adopting the deeper convolution neural network (CNN) was proposed to obtain high-quality reconstruction result and achieves state-of-the-art performance \cite{kim2016accurate,zhang2018image,zhang2018residual}. Among these methods, the ResNet \cite{resnet} based SR model \cite{zhang2018residual} shows the state-of-the-art performance, demonstrating that a stronger baseline network should be able to improve the performace as well. However, it is still hard to reconstruct the fine spatial details (such as textures) solely based on VDSR or its improved versions. 

In \cite{srgan}, the authors first adopt generative adversarial networks (GAN) to synthesize the high-quality fine details, though these reconstructed details may slightly differ from the original ones, and this technique is known as SRGAN \cite{srgan}. 
An enhanced version of SRGAN, termed as ESRGAN, is proposed to further improve the visual quality of the reconstructed image based on several learning tricks and Residual-in-Residual Dense Block (RRDB) \cite{esrgan}. 

Although the RRDB takes the advantages of both ResNet \cite{resnet} and DenseNet \cite{densenet}, some shortcomings can be further improved. In ESRGAN, the shortcut connection is established between the input data and the feature map of the last convolution layer in an RRDB. In ResNet \cite{resnet}, the shortcut is connected between every two layers, which can fully take advantage of the effective gradient information. However, the shortcut connections in RRDB is relative few, compared to ResNet. To overcome this shortcoming, we add one more shortcut in an RRDB to ensure more effective learning of the SR model, called densely connected residual module. On the other hand, several upsampling modules have been studied recently. However, it is hard to distinguish which upsampler should we use in the SR model. In this paper, instead, two upsamplers are selected to simultaneously super-resolve the feature maps of the base network. Afterward, two super-resolved feature maps will be fused by the last convolution layer to obtain the SR image in RGB channels, termed as dual reconstruction module. As shown in Fig. \ref{fig:header}, the proposed method can provide more fine details of the reconstructed image based on training set provided by \cite{aim1, aim2}. In summary, the main contribution of the proposed method is two-fold:
\begin{enumerate}
    \item We proposed a densely connected residual (DCR) network to better capture the hierarchical feature representations.                                 
    \item The proposed dual reconstruction module (DRM) can effectively super-resolve low-resolution (LR) image based on two different upsampler modules.
\end{enumerate}

The rest of this paper is organized as follows.
Some most relevant works are surveyed in Sec.2. Sec.2 presents the proposed SR model based on DCR and DRM. In Sec.3, experimental results are demonstrated. Finally, conclusions are drawn in Sec.4.

\section{Proposed Super-Resolution Method}
%%==================Figure====================
\begin{figure*}
	\centering
	\includegraphics[width=1\textwidth]{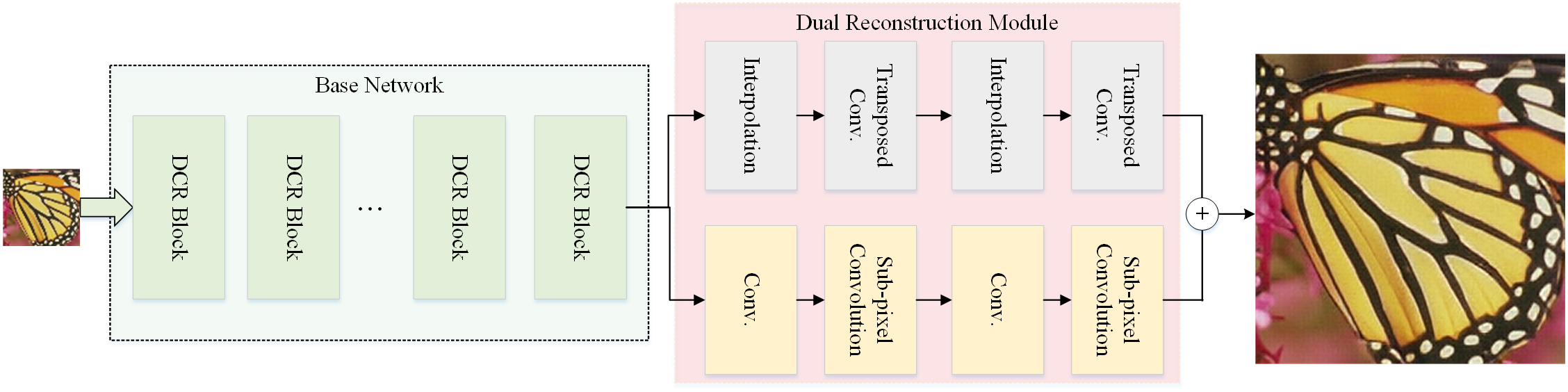}
	\caption{\small Diagram of the proposed method with the proposed dual reconstruction technique.}
	\label{fig:flowchart}
\end{figure*}
%%==================Figure====================
The overall flowchart of the proposed method is illustrated in Fig.\ref{fig:flowchart}. First, we build the base network by composing several proposed DCR blocks to extract useful features for the LR image. Second, the extracted feature maps will be upsampled by the proposed DRM simultaneously. Afterward, two super-resolved feature maps can be estimated and added to each other. Finally, the super-resolved image is obtained by the last convolution layer with $3$ channels. 

%%%==============Proposed method ==========================
Three loss functions are used in the proposed method. The first one is pixel loss, which is a common way to measure the similarity between the reconstructed image and its ground truth. Usually, the pixel loss is measured by $l1$ norm distance \cite{esrgan,srgan,zhang2018residual}, which is defined as
\begin{equation}\label{eq:loss_p}
L_p(\mathbf{x}_{HR},\mathbf{x}_{SR})=\frac{1}{N^2-1}\sum_{i=1}^{N^2} | \mathbf{x}_{SR}-\mathbf{x}_{HR} |,
\end{equation}
where $N^2$ is number of pixels of HR/SR image, $\mathbf{x}_{SR}$ and $\mathbf{x}_{LR}$ present the SR and LR images, respectively.

It is well-known that the GAN-based SR models can achieve better visual-quality of the reconstructed image, such as SRGAN \cite{srgan} and ESRGAN \cite{esrgan}. In GAN-based SR models, the  adversarial loss $L_{GAN}$ incorporated in a GAN is defined as
\begin{align}\label{eq:loss_gan}
&L_{GAN}(D,G) 
\nonumber
\\
&=E_{D} \left [ \log D(\mathbf{x}_{HR}) \right ]
+ E_{G}\left[ \log \left(1- D(G(\mathbf{x}_{LR})) \right) 
\right],
\end{align}
where $D$ and $G$ indicate the discriminator and generator, respectively, $G(\mathbf{x}_{LR})$ is the generative model used for generating the super-resolved image $\mathbf{x}_{SR}$, and $D(\mathbf{x})$ is the classification of data sample $\mathbf{x}$ being visually-qualified. 

Inspired by \cite{esrgan}, the feature distance/similarity between the super-resolved image and its ground truth is usually used to measure the perceptual loss $L_{feat}$, as follows:
\begin{align}\label{eq:loss_feat}
	&L_{feat}(\mathbf{x}_{HR}, 
	\mathbf{x}_{SR})\nonumber
	\\
	&=\frac{1}{M}\sum_{i=0}^{M-1} ||f_e(\textbf{x}_{SR})(i) - f_e(\textbf{x}_{HR})(i)||_1,
\end{align}
where $M$ is the number of feature vector and $f_e$ is the feature extractor based on pre-trained Vgg16 model \cite{vgg}.

%%%%Proposed method I=================
\subsection{Densely Connected Residual Block}
%%==================Figure====================
\begin{figure*}
\vspace{0.5cm}
	\centering
	\includegraphics[width=1\textwidth]{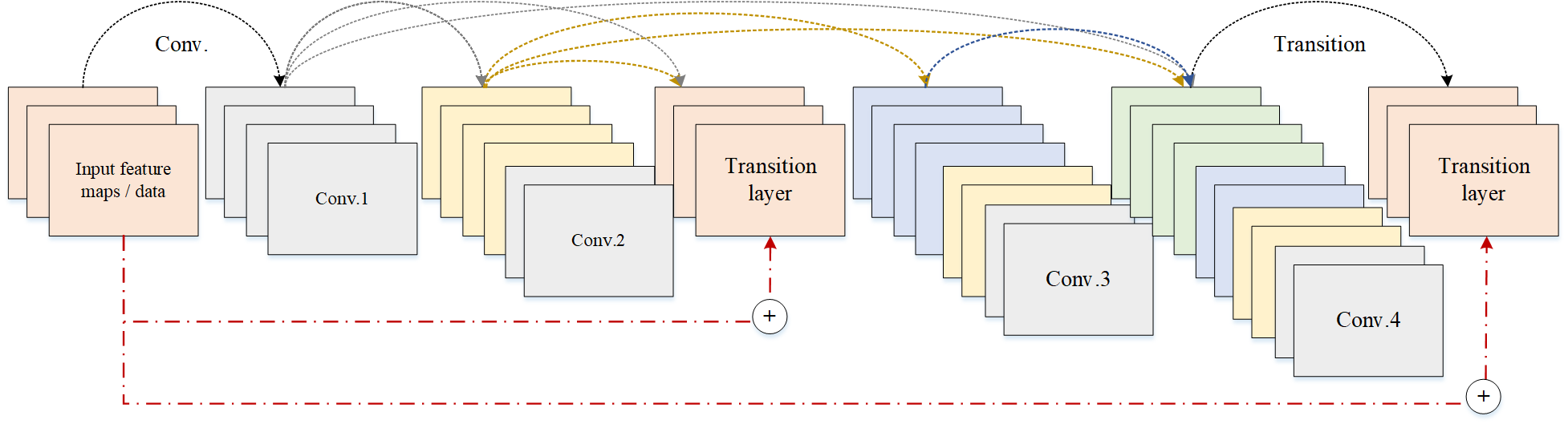}
	\caption{\small Network architecture of the proposed densely connected residual block.}
	\label{fig:DCR}
\end{figure*}
%%==================Figure====================

 Assume that the input feature map (or data) is $\textbf{x}$ with $n_c$ channels. In ESRGAN, each RRDB contains four convolutions $D=[f_{c1}, f_{c2}, f_{c3}, f_{c4}]$. The number of channels of the output feature maps at first three convolution layers are $n_g$ and last convolution layer in a dense block is $n_c$. In the first three convolution layers will be concatenated \cite{densenet} so that the number of the channels of ouputs at the third convolution layer will be $3n_g$, while the last transition layer will squeeze the input feature map to the $n_c$-channeled feature map. Then, the shortcut will be constructed $\textbf{x} + f_{c4}(f_{c3}(f_{c2}(f_{c1}(\mathbf{x})))$. To provide more gradient information and to accelerate the learning process, we propose to insert a transition layer $f_{m2}$ to squeeze the feature map of $f_{c2}(f_{c1}(\mathbf{x}))$ to the $n_c$-channeled feature map. In this way, the second shortcut between $f_m$ and $\textbf{x}$ can be constructed, as depicted in Fig. \ref{fig:DCR}. 
 
 Similarly, the base network is constructed by several DCR blocks ($3$ DCR blocks in this paper), implying that the shortcut connection can also be added between DCR blocks. Since the number of channels of the feature maps of each DCR will be the same to each other, it is easy to construct the shortcut. In this study, the bypass is established between the first and second DCRs to take back-propagate the gradient information easily, resulting the faster learning process of our SR model. 

%%%%Proposed method II=================
\subsection{Dual High-Resolution Image Reconstruction}

It is a well-known trick that the image resolution can be improved by learning the residual of the high-frequency component, instead of directly learn the SR result in RGB domain \cite{vdsr}. Also, there is a recent research trend in the remote sensing area (see, e.g., \cite{lin2017COCNMF}), aiming at fusing the high-resolution spatial details (probably extracted from some counterpart panchromatic image) with an input low-resolution image. To take the advantages from different super-resolved results (or information from LR image), we need to have different SR models first. However, it is hard to train multiple SR models due to very time-consuming as well as design a frequency decomposition function to the deep neural network. Instead, we adopt two different upsamplers including the traditional interpolation function and sub-pixel convolution operation \cite{ps} to learn different super-resolved images from the feature maps of our base network, called double reconstruction technique. 

Since sub-pixel convolution can be used to upsample the feature maps by aggregating feature responses at different channels, the information in feature maps can be fully discovered \cite{ps}. For example, the upscaling factor is $4$, we need to give $16$-channeled feature maps to upsample. Assumed that the number of the output feature map of our base network is $n_c$, a convolution layer was added to transfer the $n_c$-channeled input feature maps to $4n_c$-channeled feature maps. Then, the upsampled feature map based on sub-pixel convolution will become $n_c$ when the scaling factor is $2$. In this manner, we can ensure that the number of channels of the upsampled feature map is $n_c$. Note that we can repeat the above process to obtain the upsampled results with scaling factor $4$. Afterward, a convolution layer will be added to obtain the $n_c$-channeled feature maps. Finally, the super-resolved feature map can be obtained by concatenating two upsampled feature maps $U_1$ and $U_2$.

In summary, the details of our DRM are listed as follows:
%%Marked by jess%% This kind of fusion approach has been found effective for both image fusion or network fusion.
\begin{itemize}
	\item First upsampler $U_1$: It contains two transposed convolution layers and Bicubic Interpolation function, as depicted in the upper pipeline in Fig. \ref{fig:flowchart}.
	\item Second upsampler $U_2$: Three convolution layers and sub-pixel convolution \cite{ps} is used to upsample the input feature map of our base network, as shown in the bottom pipeline in Fig. \ref{fig:flowchart}.
\end{itemize} 

Note that the DRM technique is not used in the AIM-Real-World-Super-Resolution-Challenge-Track1 (AIMRWSR).

\subsection{Learning Tricks}

In this paper, several useful learning-tricks are adopted to have better and faster network training, as follows:
\begin{itemize}
    \item Inspired by ESRGAN \cite{esrgan}, the adversarial learning is used to fine-tune our SR model based on the GAN loss in (\ref{eq:loss_gan}), where the Vgg-16 architecture \cite{vgg} is used as our discriminator.
	\item We adopt H-Swish \cite{mobilenetv3} activation in our discriminator to accelerate the learning process.
	\item We use unbalanced learning rates for the generator and discriminator in the second model (determined by manual check).
	\item The fully connected layers are removed from the discriminator. The global averaging pooling is used instead to better capture the spatial information of feature maps \cite{fcn}.
	\item Larger learning rate at the first $10,000$ iterations and followed by relative small learning rate to train our SR model.
	\item Since the first SR model (say, SNR-aware model, SAM) is learned by (\ref{eq:loss_p}) and the fine-tuned SR model (say, visual-aware model, VAM) is learned by (\ref{eq:loss_p})(\ref{eq:loss_gan})(\ref{eq:loss_feat}), it is worth to fuse the results of VAM and SAM to obtain the better super-resolved image. Inspired by \cite{esrgan}, the final SR model can be fused by merging the parameters in tow models such that $\alpha\text{SAM}(\mathbf{x}_{LR})+(1-\alpha)\text{AIM}(\mathbf{x}_{LR})$ for some $\alpha\in[0,1]$.
\end{itemize}

\section{Experimental Results}
\subsection{Experiments Settings}
 In our solution, we adopt Python as the programming language. Since our method is based on deep learning, we use PyTorch as our backbone. We train the SR model based on a personal computer equipped with two GTX 2080Ti GPUs, Intel i9-9900 CPU, and 64 GB system memory. In this paper, two SR models are learned sequentially. The first model is based on (\ref{eq:loss_p}). In this way, it should be able to learn the SAM SR network. Afterward, adversarial learning is adopted to learn the VAM SR model based on the trained SAM SR model.
 
 In our SAM SR model, the initial learning rate is $3e-4$ with SGDR learning decay policy \cite{sgdr}. The size of the ground truth at the first $400,000$ iterations is $128\times128$, while that at the following $400,000$ iterations is $160\times160$. The total iterations is $800,000$ with Adam optimizer \cite{adam}. In the proposed VAM SR model, the initial learning rates of the generator and the discriminator are respectively $1e-3$ and $1e-4$. Since we found that the discriminator is faster to converge, we use the unbalance learning rate setting to avoid the learning difficulty. The size of the ground truth at the first $40,000$ is $160\times160$ while that at the following $40,000$ iterations is $128\times128$. The total iterations are $80,000$ with the learning rate decay by a ratio of $0.5$ for every $5,000$ iteration. The same experiment settings are adopted in ESRGAN \cite{esrgan} to have a fair comparison. Therefore, we select the trained model at iteration $80,000$ as our final ESRGAN model. 
 
 The proposed SR model is trained based on AIMRWSR training set (source domain) \cite{aim1,aim2}. The scaling factor is $4$. Therefore, the size of LR image is $128\times128$ if the size of corresponding HR image is $512\times512$. To verify the performance of the reconstructed image during training phase, we evaluate the PSNR and SSIM based on the $6$ selected images from DIV2k\cite{div2k}. To further evaluate the performance, Set5 and Set14 in \cite{LapSRN} are used as the test set.

\subsection{Objective Quality Comparison}

%%=========Results from AIM organizers===============
\begin{table}[t]
	\centering\label{tab:aimresult}
	\newcommand{\sep}{~~}
	\resizebox{1\columnwidth}{!}
	{
		\begin{tabular}{@{}ll@{\sep}|@{\sep}c@{\sep}c@{\sep}c@{\sep}c@{}}
			& Method        & $\uparrow$PSNR & $\uparrow$SSIM & $\downarrow$LPIPS  & $\downarrow$MOS\\
			\hline
			\multirow{7}{1mm}{\rotatebox{90}{\resizebox{14mm}{!}{Participants}}}
			& MadDemon                   & 22.65 & 0.48 & 0.36 & \textbf{2.22} \\
			& IPCV IITM                  & 25.15 & 0.60 & 0.66 &         2.36 \\
			& Nam                        & 25.52 & 0.63 & 0.65 &         2.46 \\
			& CVML                       & 24.59 & 0.62 & 0.61 &         2.47 \\
			& ACVLab-NPUST (ours)        & 24.77 & 0.61 & 0.60 &         2.49 \\
			& SeeCout                    & 25.30 & 0.61 & 0.74 &         2.50 \\
			& Image Specific NN for RWSR & 24.31 & 0.60 & 0.69 &         2.56 \\
			\hline
			& ESRGAN Superv. & 24.22 & 0.55 & 0.30 & 1.81 \\
			\hline
			\multirow{4}{1mm}{\rotatebox{90}{\resizebox{12mm}{!}{Baselines}}}
			& Bicubic         & 25.34 & 0.61 & 0.73 & 2.43 \\
            & EDSR PT         & 25.14 & 0.60 & 0.67 &  \\
            & ESRGAN PT       & 22.57 & 0.51 & 0.58 &  \\
            & ESRGAN FT-Src   & 24.54 & 0.56 & 0.45 &  \\
			\hline
		\end{tabular}
	}
	\vspace*{0mm}
	\caption{Challenge results for Track 1: Source domain on the final test set \cite{aim1,aim2}}
	\label{tab:tr1-test-results}
	\vspace*{0mm}
\end{table}
%%=========Results from AIM organizers===============
%%==================Figure====================
\begin{figure*}
	\centering
	\includegraphics[width=0.75\textwidth]{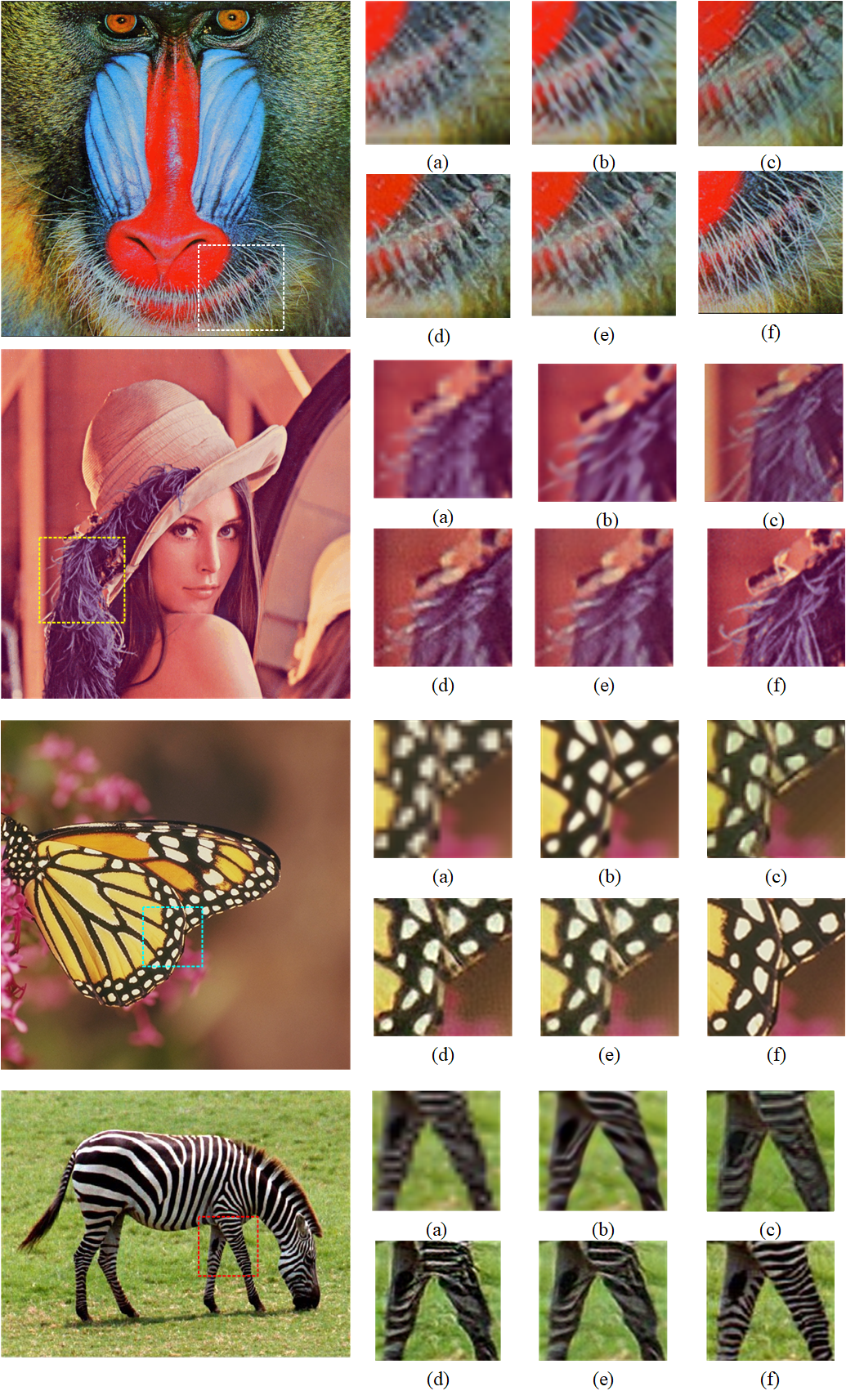}
	\caption{\small Subjective visual quality comparison of the proposed method and the baselines: (a)--(e) the reconstructed HR faces using (a) bicubic interpolation, (b) ESRGAN \cite{esrgan}, (c) SAM-SR, (d) VAM-SR, (e) VAM+SAM SR, and (f) the ground-truths}
	\label{fig:vis_result}
\end{figure*}
%%==================Figure====================
In AIMRWSR Challenge, the proposed DRM is disabled because it is hard to train the proposed SR model with DRM in a limited time. Table \ref{tab:aimresult} shows the performance comparison between the proposed method (named ACVLab-NPUST) and other methods proposed by other participants \cite{aim1, aim2}. We show good performance in terms of PSNR/SSIM/LPIPS. Since the test set did not release yet, we compare the proposed method with enabled DRM with other methods based on Set5 and Set14. Table \ref{tab:result2} shows the performance comparison between the proposed method and other SR models for Set5 and Set14. Our proposed method achieves better performance in terms of SSIM and PSNR, compared to ESRGAN \cite{esrgan}. In our VAM-SR model, the objective quality is lower than ESGAN because the GAN-based SR models tend to super-resolved the HR image with fine details, whereas the synthesized details may differ from that in the ground truth. Note that the result from ESRGAN is generated by fusing the ESRGAN and its pre-trained SRResNet model with $\alpha=0.8$ \cite{esrgan}.

\begin{table} %%% 
  \caption{Performance comparison among the different SR methods evaluated on Set5 and Set14.}
  \label{tab:result2}
  \centering
  \begin{tabular}{c|cc|cc}
    \hline
    \multirow{2}{*}{Method} &
			\multicolumn{2}{p{60pt}|}{Set5} &\multicolumn{2}{p{60pt}}{Set14}\\
			\cline{2-5}
			& PSNR & SSIM & PSNR & SSIM\\
    
    \hline
    ESRGAN \cite{esrgan}	    & 27.13 & 0.648 & 24.62 & 0.589 \\
    SAM-SR (ours)         	    & 27.64 & 0.648 & 24.85 & 0.596  \\
    VAM-SR (ours)       	    & 27.85 & 0.648 & 25.25 & 0.607 \\
    SAM+VAM (ours)              & \textbf{28.03} & \textbf{0.649} & \textbf{25.60} & \textbf{0.616}  \\
    \hline
\end{tabular}
\end{table}

\subsection{Subjective Quality Comparison}

%%==================Figure====================
\begin{figure}
	\centering
	\includegraphics[width=0.45\textwidth]{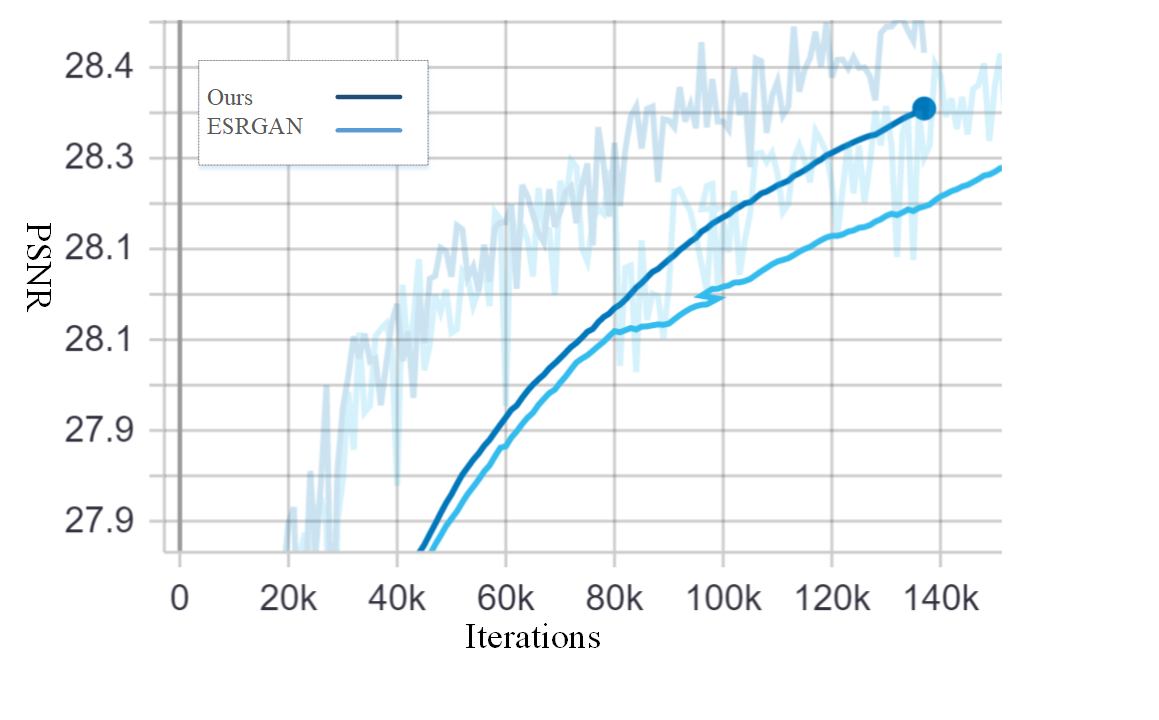}
	\caption{\small The PSNR curve of the Set5 \cite{LapSRN} measured by the ground truth and the super-resolved image during training phase.}
	\label{fig:curve}
\end{figure}
%%==================Figure====================
\begin{table} %%% 
  \caption{Ablation study on the proposed method for Set5 and Set14.}
  \label{tab:result3}
  \centering
  \begin{tabular}{c|cc|cc}
    \hline
    \multirow{2}{*}{Method} &
			\multicolumn{2}{p{60pt}|}{Set5} &\multicolumn{2}{p{60pt}}{Set14}\\
			\cline{2-5}
			& PSNR & SSIM & PSNR & SSIM\\
    
    \hline
    %    ESRGAN \cite{esrgan}	    & 27.13 & 0.648 & 24.62 & 0.589 \\

    Baseline-I (SAM)         	& 27.61 & 0.640 & 24.89 & 0.600  \\
    Baseline-I (VAM)       	    & 27.32 & 0.639 & 24.32 & 0.590 \\
    Baseline-I (VAM+SAM)        & 27.61 & 0.641 & 25.30 & 0.607 \\
    Baseline-II (SAM)         	& 27.57 & 0.641 & 25.52 & 0.612  \\
    Baseline-II (VAM)       	& 27.21 & 0.639 & 24.62 & 0.589 \\
    Baseline-II (VAM+SAM)  	    & 27.82 & 0.640 & 25.47 & 0.611 \\
    SAM-SR (ours)         	    & 27.64 & 0.648 & 24.85 & 0.596  \\
    VAM-SR (ours)       	    & 27.85 & 0.648 & 25.25 & 0.607 \\
    SAM+VAM (ours)              & \textbf{28.03} & \textbf{0.649} & \textbf{25.60} & \textbf{0.616}  \\
    \hline
\end{tabular}
\end{table}
Figure \ref{fig:vis_result} presents the subjective quality comparison between the proposed method and ESRGAN. It is clear that the proposed method (SAM+VAM) can provide fine details, while the ESRGAN also generates fine synthesized textures. However, the details in the reconstructed image based on our method is more realistic, compared to ESRGAN. It demonstrates that the DRM module indeed can be used to improve the visual quality of the reconstructed image further. In the limited iterations, ESRGAN can not learn the good visually-pleased results from LR images. Although the textures in the reconstructed image based on SAM-SR model cannot be fully recovered, the super-resolved results will be closer to the ground truth. 

\subsection{Ablation Study}
In this subsection, the DCR and DRM are removed respectively to form the Baseline-I and Baseline-II. Two Baselines are trained on the same training set provided by AIMRWSR challenge \cite{aim1, aim2}. It is verified that the proposed DCR and DRM can be used to improve the visual and objective quality of the reconstructed image. It also shows that the DRM can fuse different HR details within the network, which may exist some potential cues to synthesize the better high-quality SR images.  

On the other hand, we also show the PSNR curve for the Set5 during the training phase in Fig. \ref{fig:curve}. It is clear that the proposed SAM-SR model is relatively faster to converge, compared to the default number the of iterations in ESRGAN \cite{esrgan} (say, $1,000,000$ iterations), which indicates that the more shortcut connections should be able to accelerate the learning process as well as slightly improve the performance of the super-resolution task. In the proposed SAM-SR model, the PSNR value can be approximately improved to $28$dB at $550,000$ iteration, which is significantly smaller than the default iterations used in ESRGAN \cite{esrgan}. We also evaluate our method for the validation set in AIMRWSR challenge \cite{aim1} based on the online validation server. The PSNR values of the reconstructed images of the proposed method and Baseline-I (i.e., the submitted version in the AIMRWSR challenge for final test set) are $25.60$dB and $25.49$dB, implying that our DRM can further improve the performance. 

\section{Conclusions}
In this paper, we have proposed a novel single image super-resolution model based on the densely connected residual (DCR) blocks and dual reconstruction module. 
The proposed DCR architecture effectively accelerate the learning process because one more shortcut connections between convolution layers and DCR blocks allow more effective back-propagation of the gradient information. Furthermore, our dual reconstruction module fuses counterpart super-resolved images directly from our base network, yielding a better synthesized high-resolution image both quantitatively and qualitatively. Experimental results have demonstrated the effectiveness and efficiency of the proposed super-resolution model. 

\section*{Acknowledgment}
This study was supported in part by the Ministry of Science and Technology, Taiwan, under Grants MOST 108-2634-F-007-009, 107-2218-E-020-002-MY3, and 108-2218-E-006 -052.

{\small
\bibliographystyle{ieee_fullname}
\bibliography{egbib}
}

\end{document}